\documentclass[reprint,aps,prd,superscriptaddress,showkeys,showpacs]{revtex4-1}
\usepackage{epsfig,amsmath,natbib}

\usepackage{aas_macros}
\usepackage{amssymb}
\usepackage{amsmath}
\usepackage{dsfont}
\usepackage{hyperref}
\usepackage{color}
\usepackage{pbox}
\usepackage{enumerate}

\hypersetup{
	colorlinks=false,
	citecolor=green
}

\begin{document}

\title{Non-Gaussian information from weak lensing data via deep learning}

\author{Arushi Gupta}
\email{ag3309@columbia.edu}
\affiliation{Department of Computer Science, Columbia University, New York, NY 10027, USA}
\author{Jos\'e Manuel Zorrilla Matilla}
\email{jzorrilla@astro.columbia.edu}
\affiliation{Department of Astronomy, Columbia University, New York, NY 10027, USA}
\author{Daniel Hsu}
\affiliation{Department of Computer Science, Columbia University, New York, NY 10027, USA}
\author{Zolt\'an Haiman}
\affiliation{Department of Astronomy, Columbia University, New York, NY 10027, USA}

\date{\today}

\begin{abstract}
  Weak lensing maps contain information beyond two-point statistics on small scales.  Much recent work has tried to extract this information through a range of different observables or via nonlinear transformations of the lensing field.  Here we train and apply a 2D convolutional neural network to simulated noiseless lensing maps covering 96 different cosmological models over a range of \{$\Omega_m,\sigma_8$\}. Using the area of the confidence contour in the  \{$\Omega_m,\sigma_8$\} plane as a figure-of-merit, derived from simulated convergence maps smoothed on a scale of 1.0 arcmin, we show that the neural network yields $\approx 5 \times$ tighter constraints than the power spectrum, and $\approx 4 \times$ tighter than the lensing peaks. Such gains illustrate the extent to which weak lensing data encode cosmological information not accessible to the power spectrum or even other, non-Gaussian statistics such as lensing peaks.
\end{abstract}

\maketitle

\section{Introduction}\label{ml_introduction}
The analysis of multiple probes, including the Cosmic Microwave Background (CMB) and large scale structure (LSS), have yielded very precise estimates for the parameters that define the standard cosmological model, $\Lambda$-CDM \cite{Planck15,Anderson14}.  Early fluctuations in the CMB evolved through gravitational instability and formed the structures we observe in the late universe. The evolution of the matter distribution in the universe encodes rich cosmological information that can be mined to test the standard model and constrain the possible values for its defining parameters.

Over 80\% of the matter in the universe is non-baryonic Dark Matter (DM), detectable through its gravitational effects. It contributes to gravitational lensing, distorting the shapes of background galaxies to an extent that is usually too small to be directly observed. Weak gravitational lensing (WL) can, nonetheless, be measured statistically through the correlation in the shapes of galaxies \cite{Refregier03, Kilbinger15}. The lensed galaxies' redshifts allow the reconstruction of the matter density field's evolution \cite{Hu02}, making WL one of the most promising cosmological probes. Lensing measurements and their analysis in a cosmological context are an essential part of experiments such as CFHTLenS \footnote{\url{http://www.cfhtlens.org/}}, KiDS \footnote{\url{http://kids.strw.leidenuniv.nl/index.php}}, the Dark Energy Survey (DES \footnote{\url{http://www.darkenergysurvey.org}}) or HSC \footnote{\url{http://hsc.mtk.nao.ac.jp/ssp/}}, and will be included in even wider ($\approx 10 \times$ larger) surveys (Large Synoptic Survey Telescope, LSST \footnote{\url{http://www.lsst.org}}, the Euclid mission \footnote{\url{http://sci.esa.int/euclid/}} and the Wide Field Infrared Survey Telescope, WFIRST \footnote{\url{http://wfirst.gsfc.nasa.gov}}).

The large volume of upcoming datasets raises the question of how to extract all the cosmological information encoded in them. Non-linear gravitational collapse distorts the Gaussian character of the initial fluctuations. Thus, two-point statistics are insufficient to characterize weak lensing data and additional descriptors have been considered to extract additional information \cite{Weinberg13}. An alternative approach is to transform the data so that non-linearities become less important and it is easier to recover the information encoded in the transformed field (e.g. with the power spectrum). Logarithmic transformations have been proposed for the 3D matter density field \cite{Neyrinck09} and the 2D convergence \cite{Seo11}, as well as other local, Gaussianization transformations \cite{Shirasaki17}. 

Overall, non-Gaussian statistics such as lensing peaks and moments involving gradients of the convergence field are promising, since they can improve parameter errors by a factor of 2-3 compared to using only second-order statistics  \cite{Kratochvil10, Kratochvil12, Dietrich10, Shirasaki14, Liu15, Petri15, DES16, Martinet18}. It is not clear where the extra information lies, or if all of it is accessible \cite{Takada14}. It has been investigated and partially understood only for lensing peaks, which derive some (but not all) information from underlying collapsed DM halos \cite{Yang11, Zorrilla16, Liu16}. This halo-peak connection has inspired the development of approximate analytic models for peak counts \cite{Fan+2010, Lin15a}.

All these statistics compress the information in the original dataset, typically a map representing a noisy estimate of the projected matter density field, into a low-dimensional descriptor that can be used to infer the parameters that determine how the data was generated. An alternative approach is to use deep learning techniques, which have proven successful in a wide range of areas \cite{LeCun15} to infer cosmological parameters directly from the uncompressed raw data. 

Artificial neural networks (NNs) are pattern recognition algorithms, in which a series of processing nodes, capable of performing simple operations, are connected to each other in a network. The nodes of a NN are typically arranged in layers, with nodes in one layer connected to those in the next. Information is fed to the NN through the input layer, its outcome comes from the output layer, and all intermediate steps are called ``hidden" layers. The strength of the connections is stored in a series of weights that can be adjusted to match a given output; this process is called "learning". This quality allows the use of NNs for forecasting and inference. While we do not have a full understanding on what drives NNs predictive power \cite{Lin16}, they have been successfully used in Astronomy, from source detection and classification, to light curve analyses and even adaptive optics control (see reviews on NNs in astronomy in \cite{Miller93, Tagliaferri03}).

Convolutional Neural Networks (CNNs) are particularly well suited to work on datasets with spatial information, such as images, since the connection of their convolution layers' nodes to subsets of the data take advantage of the high correlation of nearby points imprinted by the locality of physical processes. Recently they have been used to infer cosmological parameters from the 3D matter density field \cite{Ravanbakhsh16}, and have been found to outperform constraints estimated from its power spectrum. Weak lensing provides (in principle) an unbiased map of the projected matter distribution. One of the aims of this study is to assess if neural networks over-perform relative to the power spectrum when analyzing 2D WL data, as they do for the 3D matter field. Similar techniques have also been used to generate data with the same statistical properties as the output of physically-motivated simulations \cite{Mustafa17, Rodriguez18}.

A similar study has recently applied convolutional neural networks to weak lensing data for inference \cite{Schmelzle17}. Our study shares the same motivation and reaches similar conclusions, but has some differences. While \cite{Schmelzle17} focused on the ability of deep learning techniques to differentiate between models along a known \{$\Omega_m,\sigma_8$\} degeneracy, $\Sigma_8 \equiv \sigma_8\left(\Omega_m/2\right)^{0.6}$ \cite{Jain97}, we focus on the parameters' constraints that can be inferred by extracting information through neural networks. To do so we trained our networks on a  set of 96 cosmological models covering a large region of the parameter space (see Fig.~\ref{ml_models} for the distribution of those models). Furthermore, we used different simulation techniques, the architecture of our network is different and we compared the neural network to a different set of observables (power spectrum and lensing peaks, instead of skewness and kurtosis). Finally, we restricted our analysis to noiseless data, leaving the analysis of the effect of shape noise for a follow-up study.

The paper is organized as follows, in \S\ref{ml_methods} we describe how we generated the data used to train and test the CNN, the architecture and training of the network, and the summary statistics used as benchmarks. In \S\ref{ml_results} we compare the performance of the CNN to that of alternative summary statistics, in terms of its predictive accuracy and the cosmological constraints that can be inferred. In \S\ref{ml_discussion} we discuss the implications of our results and we summarize our conclusions in \S\ref{ml_conclusions}. 

\section{Data}\label{ml_methods}
The goal of this paper is to assess the performance of CNNs predicting cosmological parameters from WL data. We do so comparing the network's predictions with those that can be inferred from statistics measured on the maps, as well as the credible regions that can be inferred around the predicted parameters. In this section, we describe how the WL data used was generated, the design and training of the CNN, and describe the summary statistics measured on the WL data: the power spectrum and lensing peaks.

\subsection{Mock convergence maps}\label{ml_simulations}

\begin{figure}
\begin{center}
\includegraphics[width=0.5\textwidth]{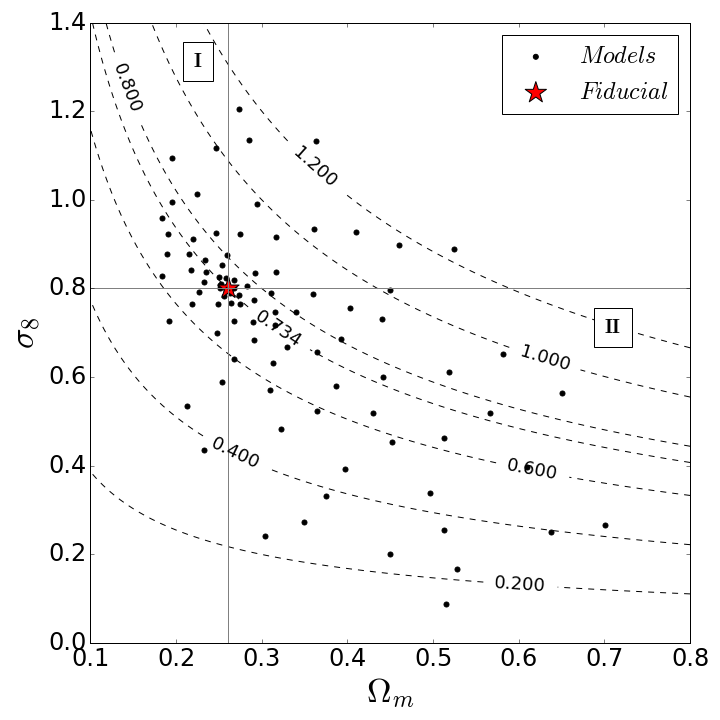}
\end{center}
\caption{\it Location of the 96 cosmological models in our dataset on the $\{\Omega_m,\sigma_8\}$ plane. The fiducial model, $\{\Omega_m=0.260,\sigma_8=0.800\}$, is marked by a red star, and grey lines delimitate the quadrants defined by the fiducial parameters. The quadrants labeled $\mathbf{I}$ and $\mathbf{II}$ are discussed in \S\ref{ml_bias}. The dashed curves show isolines for $\Sigma_8 \equiv \sigma_8\left(\frac{\Omega_m}{0.3}\right)^{0.6}$ for reference.} \label{ml_models}
\end{figure}

Our initial data set consists of mock convergence ($\kappa$) maps generated assuming 96 different values for the matter density $\Omega_m$ and the scale of the initial perturbations normalized at the late Universe, $\sigma_8$ (see Fig.~\ref{ml_models}). We adjusted the Dark Energy density to enforce flatness, $\Omega_{\operatorname{DE}}=1.0-\Omega_m$, and kept the rest of the parameters constant: baryon density ($\Omega_b=0.046$), Hubble constant ($h=0.72$), scalar spectral index ($n_s=0.96$), effective number of relativistic degrees of freedom ($n_{\operatorname{eff}}=3.04$) and neutrino masses ($m_{\nu}=0.0$). 

We singled out the cosmology with  \{$\Omega_m=0.260,\sigma_8=0.800$\} as a fiducial to compute the covariance of the observables used to assess the performance of the CNN (see \S~\ref{ml_stats}). The density of the model sampling increases towards the fiducial and shows some correlation with the direction of the $\Sigma_8$ degeneracy, $\Sigma_8 = \sigma_8 \left(\frac{\Omega_m}{0.3}\right)^{0.6}$, as can be seen in Fig.~\ref{ml_models}. We refer the reader to \cite{Zorrilla16}, where this suite of simulations was also used, and our pipeline \textsc{Lenstools} \cite{Petri16Lenstools} for a detailed description of our sampling algorithm and simulation processing. We provide a summary here for convenience.

We evolved the matter density field using the $N$-body code \textsc{GADGET2} \cite{Springel05}. For each cosmology we simulated a single volume from initial conditions computed with \textsc{CAMB} \cite{Lewis99}. The simulation boxes are cubes with a side-length of $240\,h^{-1}$Mpc, large enough to cover the maps' field of view of $3.5\times3.5\deg^2$ to a redshift of $z\approx3.0$. Each box is populated with $512^3$ Dark Matter (DM) particles, yielding a mass resolution of $\approx10^{10}M_\odot$.

We ray-traced the outputs of our simulations following the multiple lens plane algorithm \cite{Schneider92}. It has been shown that while the Born approximation is sufficient for an accurate estimation of the power spectrum even in the largest planned future WL surveys, full ray-tracing is necessary to avoid biased estimations for the counts of lensing peaks and higher order statistics \cite{Petri17}. The value of $\kappa$ for each of our maps' pixels is derived from the deflection experienced by a light ray as it crosses a series of lens planes stacked to form its past light-cone. For this study, we considered all the lensed galaxies located at a single fixed redshift of $z=1.0$. Each resulting map has $1024\times1024$ pixels, and was sliced in 16 smaller patches of $256\times256$ pixels each to speed up the neural network's training (\S\ref{ml_nn}).

Each lens plane was generated from the snapshot corresponding to its redshift by cutting a $80 h^{-1}$Mpc slab along one of its axes, estimating the matter density on a $4096\times4096$ grid, and solving the Poisson equation in 2D for the gravitational potential. By cutting different slabs, combining different planes at each redshifts, and randomly translating and rotating them, we ultimately generated 512 independent $\kappa$ maps from a single simulation box for each cosmology. Through this recycling process, it is possible to generate up to $\approx 10^4$ independent realizations of the convergence field from a single N-body simulation \cite{Petri16a}. The resulting un-smoothed, noiseless convergence maps, is analogous to a 2D version of the dataset used in \cite{Ravanbakhsh16}. 

\subsection{Neural network training and architecture}\label{ml_nn}
Neural networks consist of interconnected nodes (or neurons), arranged in layers. Each neuron transforms a linear combination of its inputs through an ``activation" function, $f(\mathbf{W}  \mathbf{x})$, where $\mathbf{W}$ is a matrix of weights and $\mathbf{x}$ a vector of inputs (in our case, the latter contains the values of the convergence in the pixelized 2D lensing map). The inputs can come from other neurons in the network, or from external data. The activation function is usually non-linear (e.g. a sigmoid function). The weights used to linearly combine the inputs can be adjusted to minimize a loss function, in a process that is called ``training" or ``learning". Some of the layers in the neural network used for this study convolve their input data with a kernel whose values are fitted during training. The resulting ``convolutional neural network" takes advantage of the correlations between neighboring pixels and has been shown to yield good results when analyzing natural images. 

Each ``labeled example" the network is exposed to is a 1024 $\times$ 1024 map coupled with the $\{\Omega_m, \sigma_8\}$ ``label" that corresponds to the cosmology used to generate that map. From each such example, we created 16 ``labeled examples" by slicing the map into smaller, 256 $\times$ 256 maps. And these are the maps used as input for the neural net. This operation reduced the number of nodes in the CNN and, consequently, its training time. We do not expect the performance of the network to be adversely affected, because of the limited constraining power of the modes that are small enough to be captured by the full maps but not their slices, i.e. spherical harmonic indices in the range $\ell \in[100,400]$ (see, e.g. ref~\cite{FangHaiman2007}, for a demonstration that most of the information is on smaller scales). The prediction for each 1024 $\times$ 1024 map is the mean of the predictions for the 16, 256 $\times$ 256 maps that were sliced from the original, bigger map. Our whole dataset amounts to 96 different cosmological models, each having 512, 1024 $\times$ 1024 independent maps. We trained the neural networks using 70\% of our data, and set aside the remaining 30\% to test their performance. 

\begin{figure*}
\begin{center}
\includegraphics[width=1.0\textwidth]{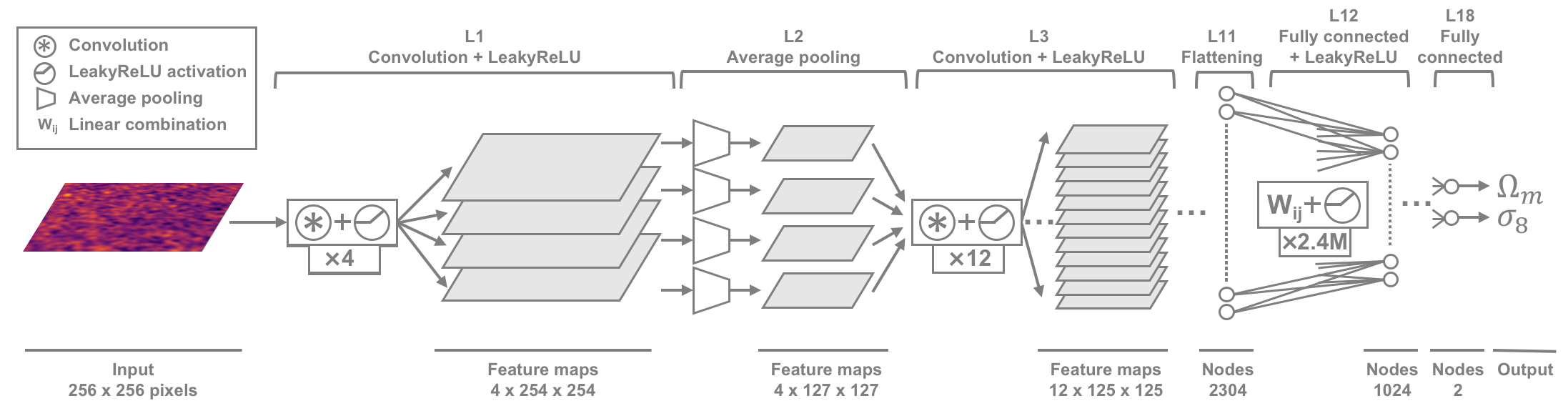}
\end{center}
\caption{\it Schematic representation of the convolutional neural network (CNN) used in this study. The network consists of a series of convolutional and (average) pooling layers. Layers increase their ``logical" dimension (depth), while reducing their ``spatial" dimensions (width and height). Once the spatial dimension has been reduced to unity (flattening), a series of fully connected layers further reduces the number of nodes to two, the required number of outputs. The activation function for the neurons is a leaky rectified linear unit. For clarity, only a few layers are displayed.} \label{ml_architecture}
\end{figure*}

The architecture of the CNN was inspired by that used in \cite{Ravanbakhsh16}. We sketch the architecture in  Fig.~\ref{ml_architecture} and summarize its elements in Table~\ref{ml_architecture_table}. The network is a combination of convolutions (transformed by a non-linear ``activation" function) and pooling layers that reduce the spatial dimensionality of the output, followed by fully connected layers in charge of the high-level logic. For the convolutional layers, we chose a $3\times3$ kernel for speed. Each convolution layer applies more than 1 filter to its input in sub-layers. The weights (filter values) are the same for all the neurons within a sub-layer. This parameter-sharing reduces the number of weights to fit during training and is a reasonable choice given the data's translational and rotational symmetries.

The first layer convolves any input map with 4 different filters and applies the activation function to the resulting 4 feature maps. Each filter is defined by 10 parameters (9 determine the convolution kernel plus an overall additive bias). In total, 40 weights need to be adjusted during training for the first layer. The second layer downsamples the feature maps from the first layer substituting $2\times2$ consecutive pixels by their mean (``average pooling"). The third and fourth layers are convolutional layers, and each applies 12 different kernels to all incoming feature maps, including all depth levels from the previous layer. While the convolution is a linear operation, the application of the activation function breaks the linearity. The number of tunable weights grows with each layer as new kernels are added. Another average pooling layer (layer 5) is followed by two sets of convolution + average pooling (layers 6-9). 

At each layer, we can consider the neurons arranged along 3 dimensions, 2 that follow the spatial dimensions of the feature maps fed into the layer (width and height) and another that grows with the number of filters used to process the layer's input (depth). As information flows through the network, the spatial dimensions of the feature maps shrink and the depth of nodes processing those maps grows. The convolutional layers 6 and 8 do not apply an activation function to their output. Another average pooling (layer 10), followed by a flattening layer (layer 11) reduce the spatial dimensionality to unity, with a depth of 2304. 

A series of fully connected layers (layers 12, 14 and 16) are followed by dropout layers (layers 13, 15 and 17) that shrink the depth of the output. The final fully connected layer (layer 18) outputs the estimated values for $\Omega_m$ and $\sigma_8$, which are compared with their true value through the loss function to adjust the weights in the network through back-propagation.

\begin{table*}
\caption{
\label{ml_architecture_table} 
\it Summary of the neural network's architecture. Convolutional layers increase the depth of the network by applying different filters (sub-layers) to the same input. The number of neurons in a layer is determined by the dimension of its output. The number of weights for a convolutional layer is given by $F_{out}\left(F_{in}\times9+1\right)$, where $F_{out}$ is the number of feature maps that the layer outputs and $F_{in}$ the number of feature maps the layer is fed with. A fully connected layer is defined by $\left(N_{in} + 1 \right) \times N_{out}$ weights, where $N_{in}$ is the number of nodes in the previous layer and $N_{out}$ the number of nodes in the fully connected layer.
}
\begin{ruledtabular}
\begin{tabular}{ccccc}
Layer	&	Type		&	Sub-layers	&	Output dimension	&	Weights\\
\hline
1	&	Convolution + LeakyReLU		&	4	&	$4\times254\times254$	&	40\\
2	&	Average pooling			&	1	&	$4\times127\times127$	&	0\\
3	&	Convolution + LeakyReLU		&	12	&	$12\times125\times125$	&	444\\
4	&	Convolution + LeakyReLU		&	12	&	$12\times125\times125$	&	1308\\
5	&	Average pooling			&	1	&	$12\times61\times61$	&	0\\
6	&	Convolution				&	32	&	$32\times59\times59	$	&	3488\\
7	&	Average pooling			&	1	&	$32\times29\times29	$	&	0\\
8	&	Convolution				&	64	&	$64\times27\times27	$	&	18496\\
9	&	Average pooling			&	1	&	$64\times13\times13	$	&	0\\
10	&	Average pooling			&	1	&	$64\times6\times6$		&	0\\
11	&	Flattening					&	1	&	2304					&	0\\
12	&	Fully connected	 + LeakyReLU	&	1	&	1024					&	2360320\\
13	&	Dropout					&	1	&	1024					&	0\\
14	&	Fully connected + LeakyReLU	&	1	&	256					&	262400\\
15	&	Dropout					&	1	&	256					&	0\\
16	&	Fully connected + LeakyReLU	&	1	&	10					&	2570\\
17	&	Dropout					&	1	&	10					&	0\\
18	&	Fully connected				&	1	&	2					&	22\\
\hline
Total	&							&		&						&	2649088
\end{tabular}
\end{ruledtabular}
\end{table*}

The total number of parameters to be fitted during training is $\approx2.6\cdot10^6$, a large number but very small compared with the total number of pixels in the training data set ($\approx 3.6 \times 10^{10}$).

The adopted activation function is the ``leaky rectified linear unit" (LeakyReLU), with a leak parameter of 0.03, within the range suggested in \cite{Maas13}:
\begin{eqnarray}
f(x) = \left\{ 
	\begin{array}{ll}
		x & \mbox{if } x \geq 0 \\
		0.03\,x & \mbox{if } x < 0
	\end{array}
	\right.
\end{eqnarray}
This functions helps mitigate the ``dying" ReLU problem, in which a neuron gets stuck in a region of zero gradient \cite{Nair10}. To prevent overfitting, we enforced regularization applying ``dropout"  at the fully connected layers: the output of any neuron was ignored with a $50\%$ chance \cite{Hinton12}. This process took part only during training, and the output from the nodes that were not dropped-out was doubled to compensate for the ignored neurons.

We used two loss functions to minimize during the training of our neural networks. The first one is the sum of the absolute error on $\Omega_m$ and $\sigma_8$, computed over batches of 32 maps each, in which the data is split for each pass of the training examples:

\begin{equation}
\sum_{\mathrm{map}\in \mathrm{batch}} \left|\sigma_8^{\text{pred}} - \sigma_8^{\text{true}}| + | \Omega_m^{\text{pred}} - \Omega_m^{\text{true}}\right|.
\label{eq:unweighted-loss-function}
\end{equation}

This is a popular choice, and converges faster than the sum of the squares of errors because its gradient does not necessarily cancel near zero. Due to the heterogeneous sampling in parameter space of our simulated models, the network is exposed to fewer examples from cosmologies in sparsely sampled regions. This can induce a bias in the predictions. To assess the impact of the non-uniform sampling on parameter constraints, we also used a weighted loss function:

\begin{equation}
\sum_{\mathrm{map}\in \mathrm{batch}} W_{cosmo}\left( \left|\sigma_8^{\text{pred}} - \sigma_8^{\text{true}}| + | \Omega_m^{\text{pred}} - \Omega_m^{\text{true}}\right|\right),
\label{eq:weighted-loss-function}
\end{equation}

where $W_{cosmo}$ is a weight inversely proportional to the sampling density at the location of a cosmological model in parameter space. Errors in predictions for maps from cosmologies in sparsely sampled regions are more severely penalized than those for maps from densely sampled regions. We show in \S\ref{ml_bias} that such a weighted loss function reduces the bias in the predictions, at the cost of a longer network training, but has only a limited impact on the parameter constraints inferred from the predictions.

The algorithm used to minimize the loss function was an Adam optimizer \cite{Kingma2014} with a learning rate of $10^{-4}$ and first and second moment exponential decay rates of 0.9 and 0.999, respectively.

We trained each network until the loss function converged, which took in most cases 5 epochs (an epoch is a pass of all the training examples in the data set). The training maps were split in batches and randomly reshuffled after each epoch. The networks' weights were recomputed after each batch, minimizing the total loss over the 32 tiles. Each batch took $40-50$ s on a NVIDIA K20 GPU with 5GB of on-board memory, at the NSF XSEDE facility~\footnote{\url{http://www.xsede.org}}. To further reinforce the rotation-invariance of the dataset, all maps were rotated $90\,\deg$ with a 50\% probability before feeding them to the network.

\subsection{Alternative descriptors}\label{ml_stats}
In order to assess the performance of the CNN, we compared the accuracy of its predictions with that achieved through analysis of summary statistics. We used two observables, the power spectrum and lensing peak counts. Both compress the information available in a given WL map in a data vector of dimension small compared with the number of pixels in the original map.

The power spectrum is defined as the Fourier transform of the two-point correlation function of $\kappa$ \cite{Kilbinger15}.

\begin{eqnarray}
\left< \kappa(\boldsymbol{\ell}) \kappa^*(\boldsymbol{\ell}')  \right> = \left(2\pi\right)^2 \delta_D\left(\boldsymbol{\ell} - \boldsymbol{\ell}' \right) P\left(\ell\right)
\end{eqnarray}

\begin{figure}
\begin{center}
\includegraphics[width=0.5\textwidth]{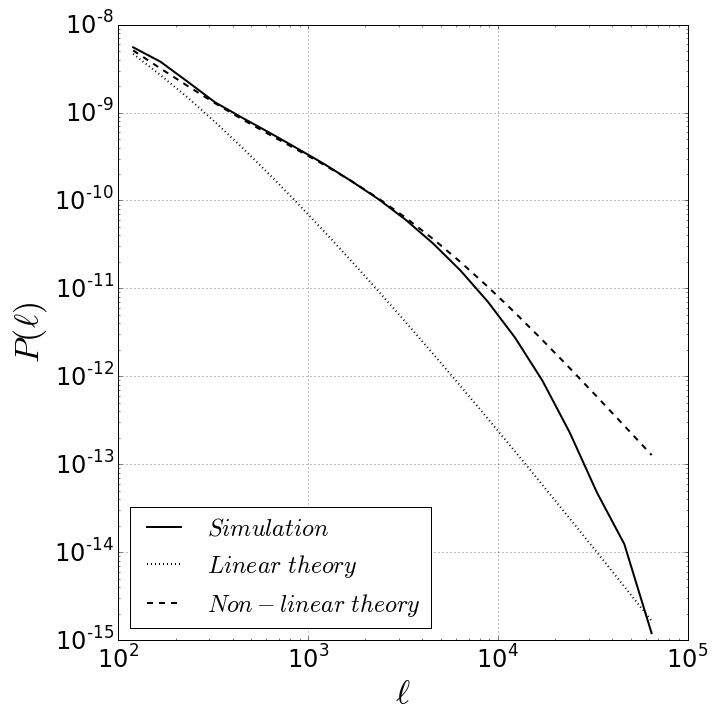}
\end{center}
\caption{\it Comparison of the average convergence power spectrum for the fiducial $\kappa$ maps with predictions from linear and non-linear theory. The theoretical curves were computed using \textsc{NICAEA} \cite{Kilbinger09}, with the revised Halofit parameters from \cite{Takahashi12}for the non-linear power spectrum.} \label{ml_ps}
\end{figure}

In the above expression $\delta_D$ is the Dirac delta function and $\boldsymbol{\ell}$ is the 2D angular wave vector. We measured the power spectrum on all 512 mock $\kappa$ maps for each of the 96 cosmological models. We evaluated the power spectra on 20 bins, logarithmically spaced in the interval $\ell \in [1\times10^2, 7.5\times10^4]$. The minimum angular scale (maximum wavenumber $\ell$) is set to prevent any loss of information at the pixel level. The finite resolution of our simulations results in deviations from theory at wavenumbers $\ell > 5\times10^3$ with a significant loss of power for $\ell \approx 10^4$, as Fig.~\ref{ml_ps} shows for the fiducial cosmology.

The power spectrum is a widely used observable in cosmology, mainly because it fully characterizes Gaussian random fields and is a well-developed analytic tool. While the initial conditions for the matter perturbations are Gaussian (or nearly so), non-linear evolution introduces significant non-Gaussianities in the matter density field at late times. 

\begin{figure}
\begin{center}
\includegraphics[width=0.5\textwidth]{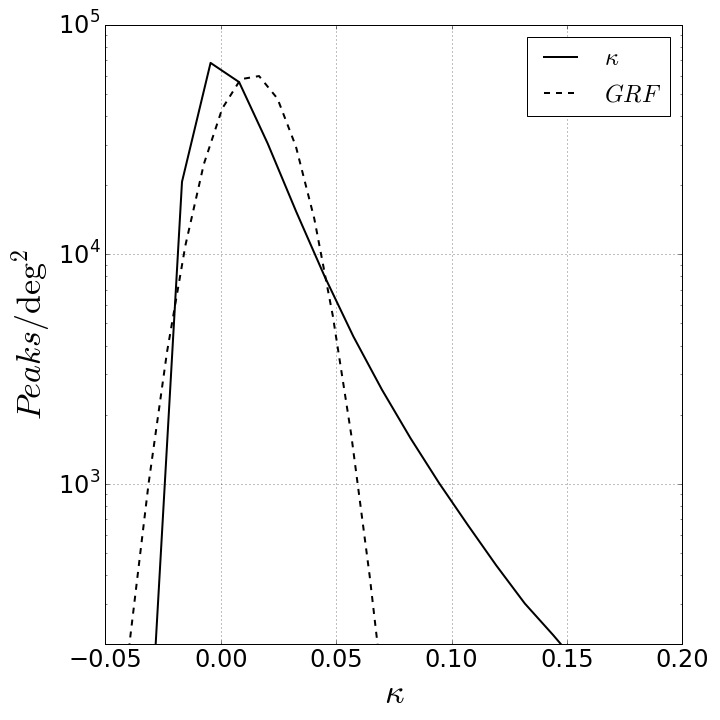}
\end{center}
\caption{\it Comparison of peak counts derived from maps generated via our ray-tracing N-body simulations, to those derived from Gaussian random fields (GRFs) with the same power spectrum.} \label{ml_pc}
\end{figure}

Lensing peaks are local maxima in the $\kappa$ field. In the absence of ellipticity noise, they probe high density regions, where non-linear effects become relevant. We chose the peaks' count as a function of their $\kappa$ value as a second observable because they are sensitive to information not captured by the power spectrum. As an illustration, we compare in Fig.~\ref{ml_pc} the average peak counts measured on the 512 mock maps generated for the fiducial cosmology to those measured over Gaussian Random Fields (GRFs) that share their power spectra with the $\kappa$ maps. That is, for each convergence map, we measured its power spectrum, built a GRF from it and measured the number of peaks in this new field. The distribution is clearly different, the peak histogram from convergence maps exhibiting a high $\kappa$ tail resulting from the non-linear growth of structures.

Peak counts yield tighter constraints than the power spectrum \cite{Kratochvil10, Dietrich10, Liu15, DES16, Kratochvil12, Shirasaki14, Petri15, Martinet18} and constitute a good benchmark for other methods which aim at extracting additional cosmological information. We counted the peaks in 20 bins, linearly spaced. We set the upper and lower limits of the bins to $[\kappa_{min}=-2.0,\kappa_{max}=12.0]$, in units of the mean $\kappa$ r.m.s.~for the fiducial maps, to fully cover the range of peaks present in the data; this corresponds to $\kappa_{min}\approx-0.03$, $\kappa_{max} \approx 0.19$ and a bin width of $\Delta \kappa \approx 0.01$.

\section{Results}\label{ml_results}
\begin{figure*}
\begin{center}
\includegraphics[width=1.0\textwidth]{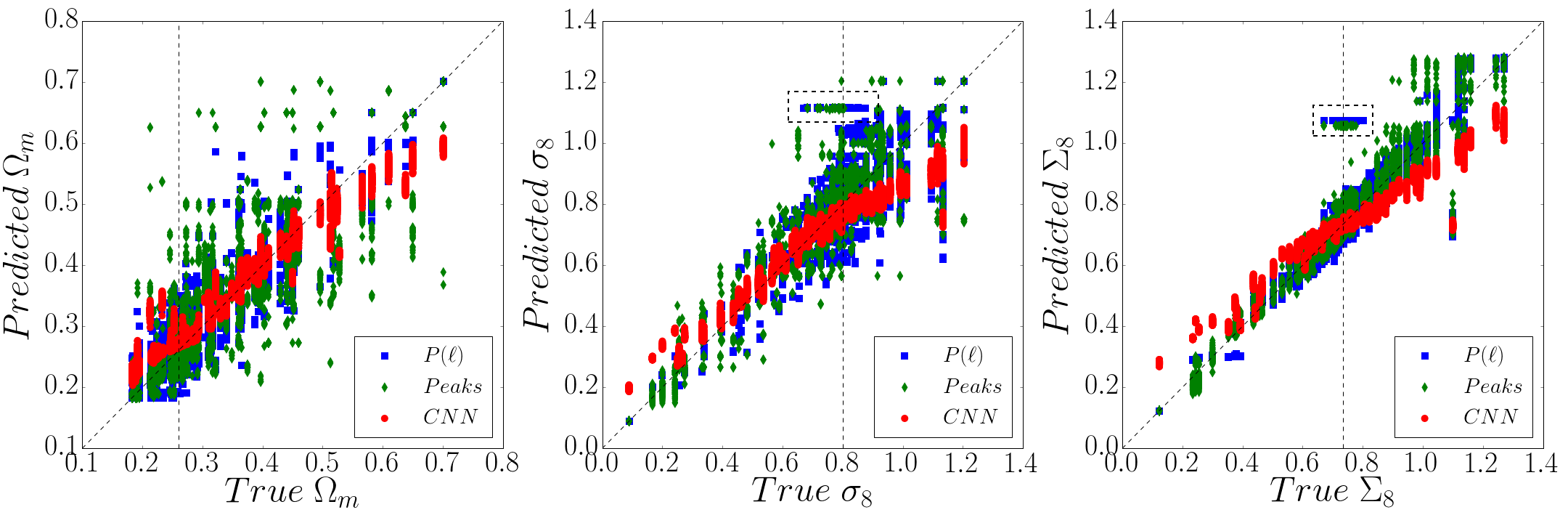}
\end{center}
\caption{\it Predictions for $\{\Omega_m,\sigma_8,\Sigma_8\}$ from un-smoothed ($\approx 0.2$ arcmin/pixel) convergence maps, compared to their true values. Each point represents a map in the test data set. Predictions from the CNN are displayed in red, from the power spectrum in blue and from peak counts in green. Vertical dashed lines indicate the true values for the fiducial cosmology, and diagonal dashed lines the unbiased $Prediction=Truth$ relationship. The dashed rectangles in the middle and right panels mark a small set of realizations of models near the fiducial cosmology; these contain anomalous structures leading to large biases (see text for discussion).  
} \label{ml_performance}
\end{figure*}

We assessed the CNN's performance in terms of the precision of their predictions for the cosmological parameters, and the constraints for those parameters for a given observation. The left and center panels of Fig.~\ref{ml_performance} display the predictions for $\Omega_m$ and $\sigma_8$ as a function of their ``ground truth", that is, the values that correspond to the cosmologies used to generate the data. The right panel shows the same comparison for the derived $\Sigma_8\equiv \sigma_8\left(\Omega_m/0.3\right)^{0.6}$ along the degeneracy between both parameters. Each point corresponds to one of the $\approx 150$ test maps available for each of the 96 cosmologies. For the neural network, the predicted $\{\Omega_m,\sigma_8\}$ for a given map are the average values for the network's output when fed the 16 tiles in which the map was sliced. For the power spectrum and peak counts, the predictions are the values that minimize $\chi^2$ for that map. We estimated $\chi^2$ for each of the 96 sampled cosmologies as:
\begin{eqnarray}
\chi^2_{ij} = \left(\mathbf{d}_i - \bar{\mathbf{d}}_j \right) \widehat{C_{fid}^{-1}} \left(\mathbf{d}_i - \bar{\mathbf{d}}_j \right)
\end{eqnarray}
where $\mathbf{d}_i$ is the data vector measured on map $i$ (binned power spectrum or peak counts), $\bar{\mathbf{d}}_j$ is the mean of the same descriptor for the model $j$ and $\widehat{C_{fid}^{-1}}$ is the precision matrix for the data vector evaluated at the fiducial model. We used all 512 available maps per model to evaluate both the mean descriptor and the precision matrix, as in any realistic scenario in which a survey provides a mass map all the simulated data would be used for inference. We corrected for the bias in the precision matrix following \cite{Hartlap07}:
\begin{eqnarray}\label{ml_CovCorr}
\widehat{C_{fid}^{-1}} = \frac{N-d-2}{N-1}C_{fid}^{-1}
\end{eqnarray}
$N$ is the number of realizations used to estimate the covariance (512)  and $d$ is the dimension of the data vector (20, the number of bins). 

The 96 $\chi^2_{ij}$ values were used to interpolate $\chi^2\left(\Omega_m,\sigma_8\right)$ and find its minimum. We used a Clough-Tocher interpolator that builds a continuously differentiable piecewise cubic surface over a non-uniform grid \cite{Alfeld84, Renka84} . The minimum was found using the downhill simplex algorithm \cite{Nelder65}. We verified that the results for the power spectrum and lensing peaks do not change when these observables are measured in a different number of bins (as long as they're more than $\approx 10$) or a different interpolator is used to find the minimum of $\chi^2\left(\Omega_m,\sigma_8\right)$.

For all cosmologies, the neural network is significantly more precise than both the power spectrum and lensing peaks: the scatter in its predictions for a given model is smaller. On average, the standard deviation of the CNN's predictions is a factor of 4-7 lower than that of the statistical descriptors, and up to $\approx 16 \times$ smaller for the fiducial (see Table~\ref{ml_perf_table}). In terms of accuracy (i.e. how close the predictions are to the ground truth), the network shows some bias that may degrade the constraints that can be inferred from the network's predictions.

\begin{table}
\caption{\label{ml_perf_table} \it Standard deviation $\left(\times 10^3\right)$ of the predictions for the parameters $\{\Omega_m,\sigma_8,\Sigma_8\}$, averaged for all the cosmological models. In parenthesis, values for the fiducial model.}
\begin{ruledtabular}
\begin{tabular}{lccc}
	&	CNN		&	Power spectrum	&	Peak counts\\
\hline
	& \multicolumn{3}{c}{Noiseless, unsmoothed}\\
\hline
$\Omega_m$	&	5.1 (2.4)		&	21.7 (21.2)	&	35.6 (13.2)\\
$\sigma_8$	&	10.1 (7.9)		&	52.7 (84.1) 	&	62.5 (63.8)\\
$\Sigma_8$	&	7.2 (4.7)		&	32.3	(73.2)	&	36.0 (59.2)\\
\end{tabular}
\end{ruledtabular}
\end{table}

We note the presence of a small set of maps from models close to the fiducial for which both the power spectrum and lensing peaks tend to over-predict $\sigma_8$ and $\Sigma_8$ as a result (the outliers on both panels correspond to the same maps). These maps form a clearly detached clump on the right-most panel of Fig.~\ref{ml_performance}, where a dashed rectangle highlights their location. They represent $\approx 4 \%$ of the maps for $\approx 28$ cosmologies not far from the fiducial model. We found through visual inspection that this over-prediction seems to be due to an anomalous number of structures projected in the field of view. Interestingly, the CNN seems to be immune  to such chance projections and classifies these maps correctly. This suggests that the neural network extracts different information from the maps than the power spectrum or lensing peaks. Alternatively, these fluctuations may be the result from cosmic variance, and the neural network may be under-weighting those effects.

For a few cosmologies, parameter predictions from the CNN converged at different values from those of neighboring models. This is noticeable on the left-most panel of Fig.~\ref{ml_performance} where a few red points show a relative over-prediction in $\Omega_m$ in the range $\Omega_m \in (0.2,0.4)$. These outliers correspond to points  in sparsely sampled areas near the boundaries of the explored parameter space. This highlights the importance of a well-sampled parameter space for the neural network to generalize accurately. In \S~\ref{ml_appendixB} we analyze the effect of sampling on the predictions and credible contours inferred from the neural network. As these outliers lie far from the fiducial cosmology, they do not alter the parameter constraints presented in this study. Furthermore, they are identifiable in the training data, and as such could be removed if needed. We did not remove any model from our data set even when it was evident from the training data that they could be outliers.

The relevant metric to compare the performance of the neural network relative to summary statistics is the probability distribution for the cosmological parameters given our data. This posterior distribution is related to the easier-to-compute probability of measuring a specific data vector given the cosmological parameters, or likelihood, by Bayes' theorem:
\begin{eqnarray}
p\left(\mathbf{p}|\mathbf{d},\mathcal{M}\right) = \frac{p\left(\mathbf{d}|\mathbf{p},\mathcal{M}\right)p\left(\mathbf{p},\mathcal{M}\right)}{p\left(\mathbf{d},\mathcal{M}\right)}, 
\end{eqnarray}
where $\mathbf{p}$ is the set of cosmological parameters, $\mathbf{d}$ a data vector and $\mathcal{M}$ the underlying model, in our case DM-only simulations of $\Lambda$CDM cosmologies. For the CNN, we define our data vector as the predicted values for the cosmological parameters, $\left(\Omega_m, \sigma_8\right)$, and for the alternative statistics, the measured binned power spectra and peak histograms described in  \S\ref{ml_stats}.

The term that multiplies the likelihood, or prior $p\left(\mathbf{p},\mathcal{M}\right)$, and that on the denominator, or evidence, are the same when using the neural network or the statistical descriptors. The reason is we are using the same convergence maps from the same sampling of the parameter space. We can drop them as a normalization factor, as well as the explicit dependence on the underlying model used to generate the $\kappa$ maps, and compare directly the likelihoods derived from the different methods. For the likelihoods, we assumed a Gaussian distribution:

\begin{equation}
p\left(\mathbf{d}|\mathbf{p}\right) \propto \exp\left[-\frac{1}{2} \left(\mathbf{d} - \bar{\mathbf{d}}(\mathbf{p}) \right)^T \widehat{C_{fid}^{-1}} \left(\mathbf{d} - \bar{\mathbf{d}}(\mathbf{p}) \right) \right],
\end{equation}

with a precision matrix $\widehat{C_{fid}^{-1}}$ evaluated at the fiducial cosmology and as an expected value for the data, $\bar{\mathbf{d}}(\mathbf{p})$, the mean value measured from the simulations for the cosmology defined by $\mathbf{p}$.

Since we use the same covariance matrix for all cosmologies, we do not need to include the normalization pre-factor. For the power spectrum, we expect the Gaussian likelihood to be accurate. Our simulated maps cover a small field of view of $3.5 \times 3.5 \,\, \deg^2$ on which the power spectrum can be measured only for  relatively high $\ell > 100$. At those scales, many modes contribute to each measurement of the power spectrum, and the central limit theorem shows that its probability distribution function should converge to a Gaussian \cite{Sato10}. For the lensing peaks and predictions from the neural network, we verified that the approximation remains valid (see \S\ref{ml_appendixA}). The alternative approach of estimating the probability density using a kernel density estimator (KDE) depends on the width of the kernel chosen, and the estimates for a large dimensional data vector such as our power spectra are noisy due to the relative limited amount of independent $\kappa$ maps realizations.

\begin{figure}
\begin{center}
\includegraphics[width=0.5\textwidth]{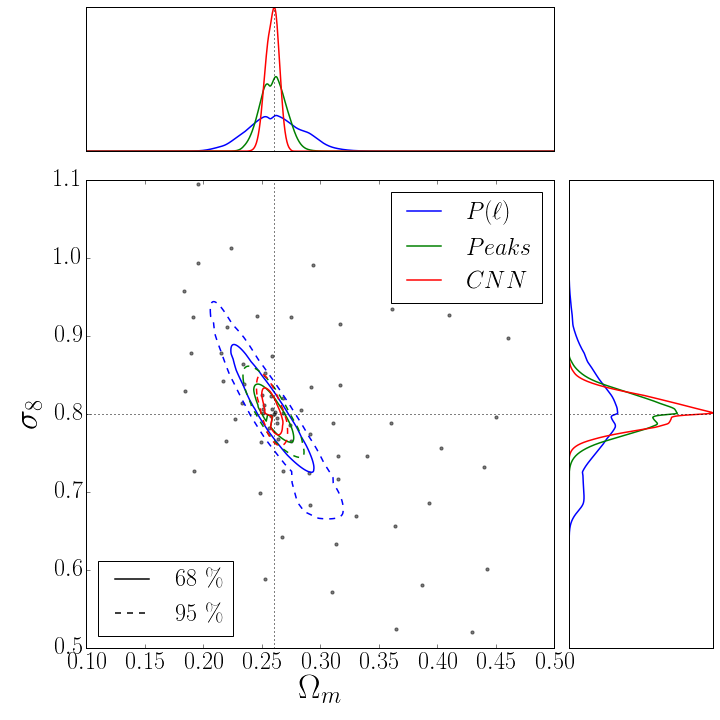}
\end{center}
\caption{\it 68\% and 95\% credible contours for un-smoothed ($\approx 0.2$ arcmin/pixel) $\kappa$ maps, derived from the power spectrum (blue), lensing peak counts (green) and neural network predictions (red). The true values for the parameters, $\{\Omega_m=0.260,\sigma_8=0.800\}$ are indicated by black dotted lines. The upper and right panels show the distribution marginalized over the other parameter.}
\label{ml_contours}
\end{figure}

To compute the likelihood, we used as data (observation) the average observable for the fiducial cosmology. For the power spectrum and lensing peaks, all 512 maps were used to estimate the means for each cosmology, and the covariance matrix for the fiducial. For the neural network, only the test maps were used ($\approx 150$ per cosmology). We display the 68\% and 95\% credible contours for the likelihoods computed for the power spectrum, lensing peaks and neural network in the central panel of Fig.~\ref{ml_contours}, and the marginalized distributions for $\Omega_m$ and $\sigma_8$ in the upper and right panels, respectively. At each point in parameter space, the expected data vector is interpolated linearly from the mean data vectors for the simulated cosmologies. Due to the choice of measurement (the predicted mean for the fiducial) all likelihoods peak at the true values for the fiducial cosmology. This is true also for the neural network. The smaller scatter in the CNN predictions translates into tighter parameter constraints, by a factor of $\approx 2$ compared with lensing peaks and $\approx 6$ compared with the power spectrum (see Table~\ref{ml_contour_table}). The neural network seems capable of extracting more information from noiseless convergence maps than alternative methods such as the power spectrum or lensing peaks.

\begin{table}
\caption{\label{ml_contour_table} \it Area of the 68\% and 95\% $\{\Omega_m,\sigma_8\}$ credible contours, relative to those obtained from the output of the neural network for un-smoothed, noiseless $\kappa$ maps.}
\begin{ruledtabular}
\begin{tabular}{lccc}
	&	CNN		&	Power spectrum	&	Peak counts\\
\hline
\hline
Area$_{68}$		&	1		&	5.9		&	1.9\\
Area$_{95}$	 	&	1		&	6.1		&	1.9\\
\end{tabular}
\end{ruledtabular}
\end{table}
%

\section{Discussion}\label{ml_discussion}
\subsection{Non-Gaussian information extracted by the neural network}
The significantly tighter constraints obtained by the CNN, shown in Fig.~\ref{ml_contours}, are encouraging and an indication that weak lensing maps encode more information than what is usually used for inference. Neural networks are capable of extracting some of it,  at least more than the power spectrum and even more than some non-Gaussian statistics such as lensing peaks. Given the large number of parameters that need to be fitted during training, there is the risk that the gain in precision comes from a form of overfitting, in the general sense of making predictions based on irrelevant information \cite{Zhang16}.

For instance, a Gaussian random field, GRF, is fully determined by its power spectrum. As a result, no other statistic or method used to extract information from it should out-perform the power spectrum. To test whether the neural network satisfies this limit, we built a collection of GRFs and used it as a new dataset to train and test the CNN's architecture. We generated the GRFs by Fourier transforming random fields with a Gaussian distribution defined by the power spectra measured over the $\kappa$ maps ray-traced from the outputs of cosmological N-body simulations. The new suite, which has a one-to-one correspondence with the original data, has no information encoded beyond the power spectrum.

\begin{figure}
\begin{center}
\includegraphics[width=0.5\textwidth]{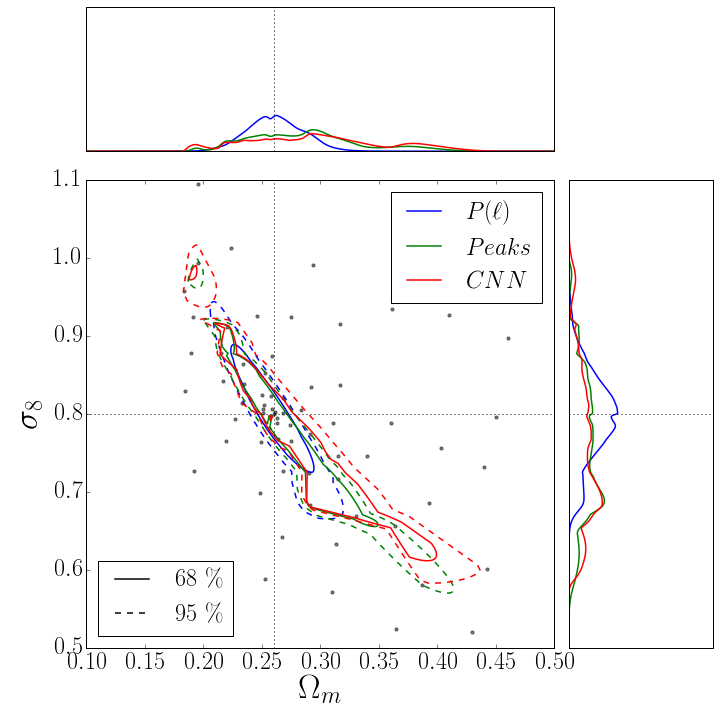}
\end{center}
\caption{\it Same as Fig.~\ref{ml_contours}, except using the Gaussian random fields, rather than the ray-tracing simulations. The network was trained with the un-weighted loss function (eq.~\ref{eq:unweighted-loss-function}).}
\label{ml_contours_grf}
\end{figure}

The 68\% and 95\% credible contours from the power spectrum, lensing peaks and the newly trained CNN, as well as the marginalized distributions for $\Omega_m$ and $\sigma_8$ are displayed in Fig.~\ref{ml_contours_grf}, which is analogous to Fig.~\ref{ml_contours} but for GRFs instead of $\kappa$ maps from N-body simulations.

As before, the likelihoods peak on the true parameter values for the fiducial and the contours appear centered around $\{\Omega_m=0.260, \sigma_8=0.800\}$. The likelihood for the power spectrum is the same as the one computed for the convergence maps. The likelihoods for lensing peaks and the neural network are different, and their contours larger than those derived from the power spectrum. In particular, the contours from lensing peaks are $1.7 \, (1.4) \times$ larger for the $68 \, (95) \%$ contours, and those from the neural network $2.6 \, (2.0) \times$ larger. This result is consistent with the absence of information beyond the power spectrum in the Gaussian Random Fields, and demonstrates that the small scatter in the parameters' predictions from the neural network trained on convergence maps is not the result of a tendency to overfitting by its architecture or other spurious effects.

\begin{figure*}
\begin{center}
\includegraphics[width=1.0\textwidth]{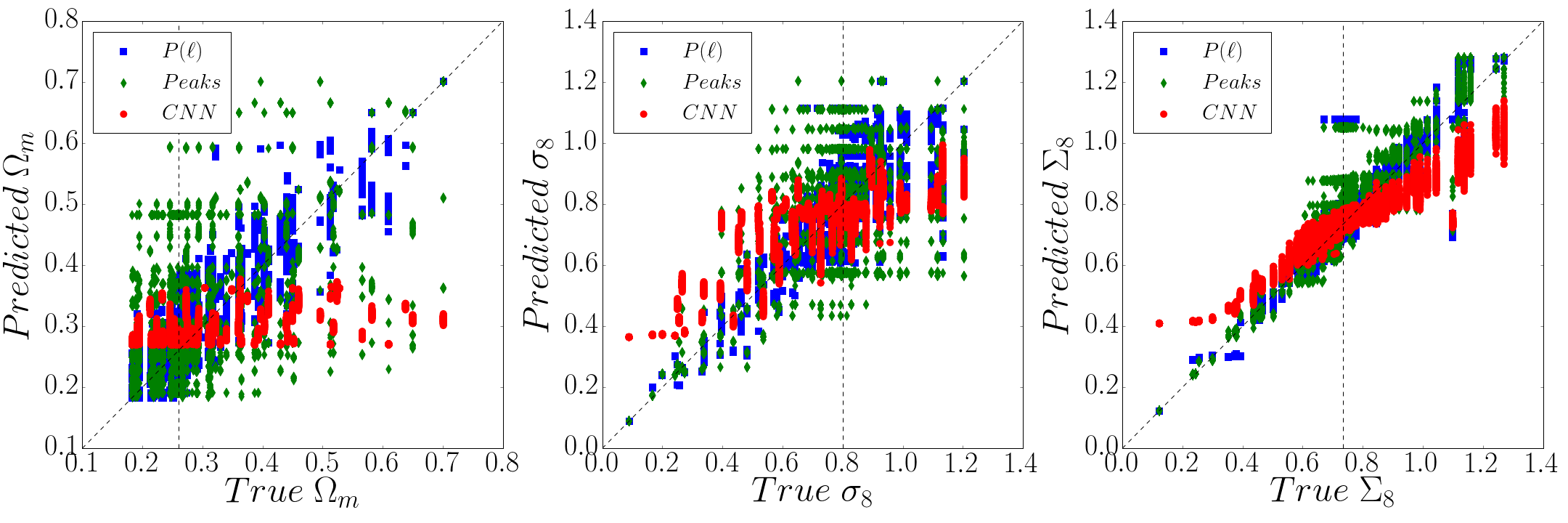}
\end{center}
\caption{\it Same as Fig.~\ref{ml_performance}, except using the Gaussian random fields, rather than the ray-tracing simulations.  The network was trained with the un-weighted loss function (eq.~\ref{eq:unweighted-loss-function}).}
\label{ml_performance_grf}
\end{figure*}

Comparing the $\{\Omega_m, \sigma_8\}$ predictions with the ground truth for the test GRFs, as done in \S~\ref{ml_results} for the $\kappa$ test maps, we see that there is both an increase in the scatter and the bias of the neural network's predictions (see Fig.~\ref{ml_performance_grf}). Both effects drive the deterioration in the parameter constraints that can be inferred from those predictions. Furthermore, the neural network seems almost insensitive to $\Omega_m$, as the predictions for all the test GRFs scatter around the median $\Omega_m$ for the 96 cosmologies. The CNN cannot easily distinguish between models with different $\Omega_m$ and defaults to the value that minimizes the loss function. The use of an unweighted loss function in this analysis may also have some influence, but the same behavior is not seen on $\sigma_8$. The power spectrum and lensing peaks are both sensitive to that parameter, indicating that they extract different information than the neural network.

\subsection{Effect of the smoothing scale on the results} 
The angular resolution of the mock convergence maps used for our analysis is $\approx 0.2$ arcmin per pixel. This high resolution is interesting from an academic perspective, but at present it is of little practical interest. Accurate shear estimates require measuring the shape of many galaxies to estimate their correlations. For instance, the upcoming LSST survey will reach an effective number of galaxies of $\approx 26 \, \text{arcmin}^{-2}$, after considering losses due to blending and masking \cite{Chang13}. This means that $\approx 1$ arcmin is characteristic of the resolution achievable by future surveys. Furthermore, at small scales ($\ell > 10^4$), baryonic physics alter the matter distribution and can bias WL observables relative to estimates from DM-only simulations \cite{Semboloni13, Osato15}. 

\begin{figure}
\begin{center}
\includegraphics[width=0.5\textwidth]{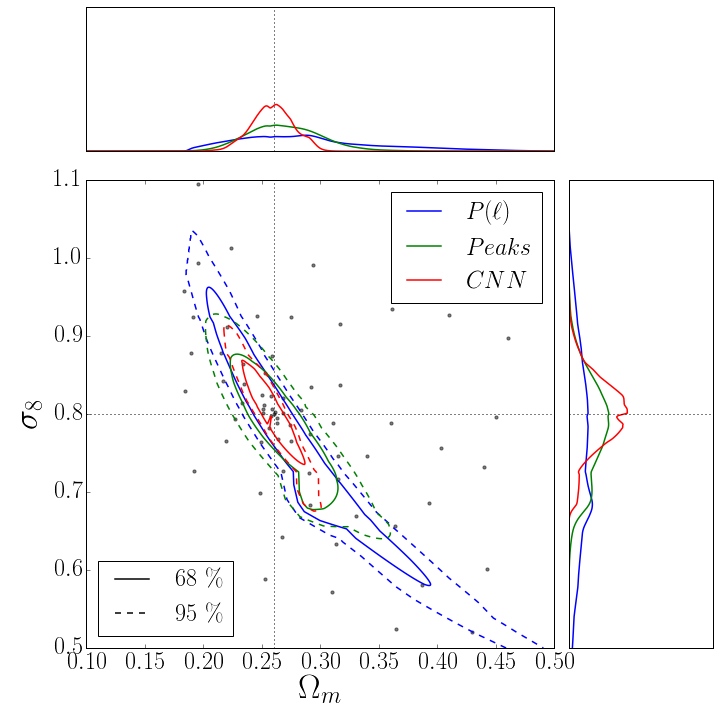}
\end{center}
\caption{\it 
  Same as Fig.~\ref{ml_contours}, except smoothing the maps from the ray-tracing simulations with a Gaussian kernel of 1 arcmin of width.  The network was trained with the un-weighted loss function (eq.~\ref{eq:unweighted-loss-function}).} \label{ml_contours_smooth}
\end{figure}

To assess whether the neural network still outperforms alternative observables on $\approx 1$ arcmin resolution data, we trained a new network with the same architecture on the $\kappa$ maps after smoothing them with a Gaussian kernel. The resulting constraints, for a smoothing scale of 1 arcmin, are displayed in Fig.~\ref{ml_contours_smooth}. 

The parameters' constrains degrade for all three methods. In principle, we would expect the non-Gaussian statistics' performance to degrade relative to the power spectrum as small scale features are smoothed away from the $\kappa$ maps. Up to 1 arcmin smoothing, the neural network keeps well its relative advantage to the power spectrum, yielding credible regions $5.6 \, (4.8) \times$ smaller at the $68\% \, (95\%)$ level. Lensing peaks are more adversely affected than the CNN by smoothing, yielding contours that are only $1.6 \, (1.5) \times$ smaller than the power spectrum. This would indicate that any additional information extracted by the neural network is not confined to very small angular scales.

The first attempt at training the neural network on smoothed data failed. To guarantee the convergence in the training process, we gradually smoothed the $\kappa$ maps in a similar way as \cite{Schmelzle17} added noise to theirs. We fed the network with maps of growing smoothing scale, starting with a kernel of 0.2 arcmin of bandwidth. Once the network reached convergence at a smoothing scale, the kernel's bandwidth was increased by 0.05 arcmin and the network re-trained.  In all cases the neural network kept its advantage (see Table \ref{ml_contour_table_smoothing}). The ratio between the areas of the credible regions derived from the power spectrum and the neural network remained roughly constant, while the same ratio for the lensing peaks and neural network increased as the capability of peaks to extract information degraded faster with larger smoothing scales.

\begin{table}
\caption{\label{ml_contour_table_smoothing} \it Area of the 68\% and 95\% $\{\Omega_m,\sigma_8\}$ credible contours, relative to those obtained from the output of the neural network, for different smoothing scales of $\kappa$ maps. The first row corresponds to the un-smoothed data.}
\begin{ruledtabular}
\begin{tabular}{ccccc}
Smoothing   &	\multicolumn{2}{c}{Power spectrum}	&\multicolumn{2}{c}{Peak counts}\\
\cline{2-3} 
\cline{4-5}
$\left[\text{arcmin}\right]$	& 68 \% 	& 95\%   	& 68\% & 95\%  \\
\hline
-		& 5.9		& 6.1 	& 1.9 	& 1.9\\
0.2		& 7.0 	& 5.9 	& 1.8 	& 1.6 \\
0.3		& 7.7 	& 7.9 	& 2.3 	& 2.7 \\
0.4		& 6.5 	& 6.4 	& 1.9 	& 2.2 \\
0.5		& 7.1 	& 6.5 	& 2.5 	& 2.5 \\
0.6		& 6.5 	& 5.7 	& 2.5 	& 2.4 \\
0.7		& 6.4 	& 5.2 	& 2.8 	& 2.4 \\
0.8		& 4.7 	& 4.1 	& 2.5 	& 2.3 \\
0.9		& 5.2 	& 4.4 	& 3.0 	& 2.8 \\
1.0		& 5.6 	& 4.8 	& 3.6		& 3.3 \\
\end{tabular}
\end{ruledtabular}
\end{table}
%

\subsection{Bias in the CNN predictions}\label{ml_bias}
\begin{figure*}
\begin{center}
\includegraphics[width=1.0\textwidth]{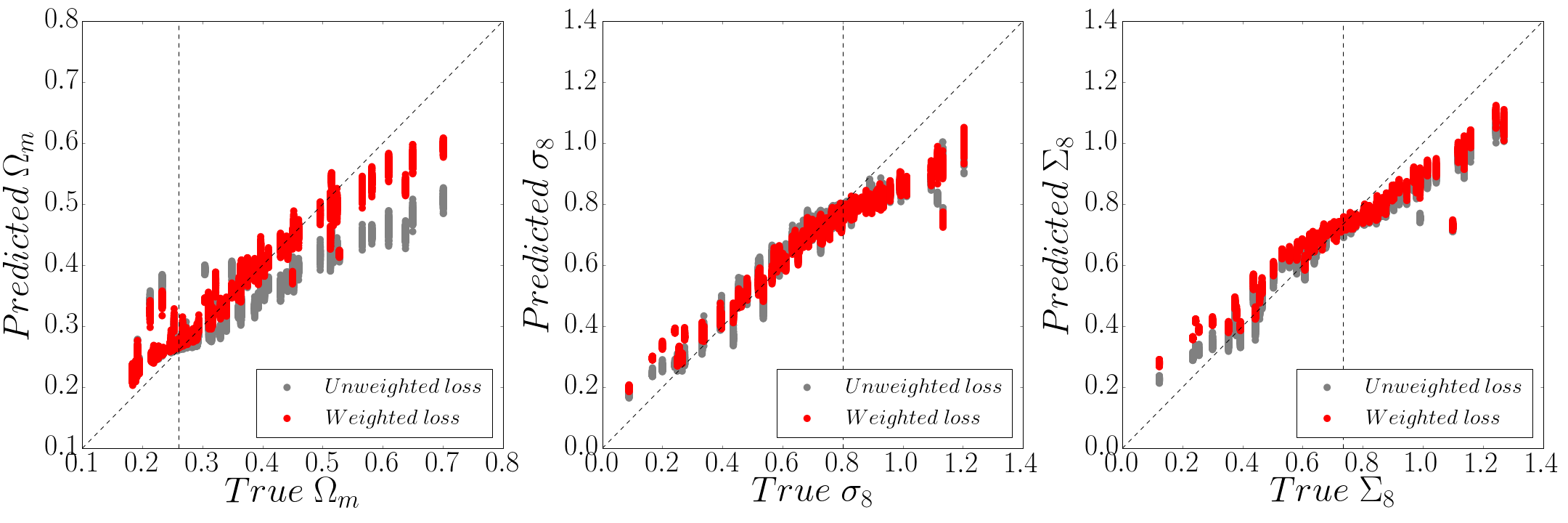}
\end{center}
\caption{\it Predictions from the CNN for $\{\Omega_m,\sigma_8,\Sigma_8\}$ from unsmoothed ($\approx 0.2$ arcmin/pixel) convergence maps, compared to their true values. Each point represents a map in the test data set. Predictions using the unweighted loss function (eq.~\ref{eq:unweighted-loss-function}) are displayed in grey, and those using a weighted loss function (eq.~\ref{eq:weighted-loss-function}), to account for the heterogeneous sampling of the parameter space, in red. Vertical dashed lines indicate the true values for the fiducial cosmology, and diagonal dashed lines the unbiased $Prediction=Truth$ relationship.} \label{ml_bias_1}
\end{figure*}

The parameter predictions from the neural network exhibit some bias (see Fig.~\ref{ml_performance}). The bias is more severe when an unweighted loss function is used, as can be seen in Fig.~\ref{ml_bias_1}. This can be due to the loss function being dominated by errors in the densely sampled regions of the parameter space.

Weighting the loss function according to the sampling density helps mitigate the bias. The effect is larger for the high-$\Omega_m$ region than for the high-$\sigma_8$ models. This can be due to the difference in sampling between both regions. The high-$\Omega_m$ region, corresponding to quadrant $\mathbf{II}$ in Fig.~\ref{ml_models} has more models further from the fiducial and with large spacing between them than the high-$\sigma_8$ region (quadrant $\mathbf{I}$).

The weights in the loss function were computed using a kernel density estimator (KDE) to estimate the sampling density in parameter space. The KDE bandwidth used was 1.0, a value that yielded a smooth estimate.

Biases in predictions from neural networks have been found in other works (e.g. \cite{Ravanbakhsh16}), so we cannot guarantee that the heterogeneous sampling of our data is the only source of the bias. Future work using a different dataset, uniformly sampled, will address this issue.

\begin{figure}
\begin{center}
\includegraphics[width=0.5\textwidth]{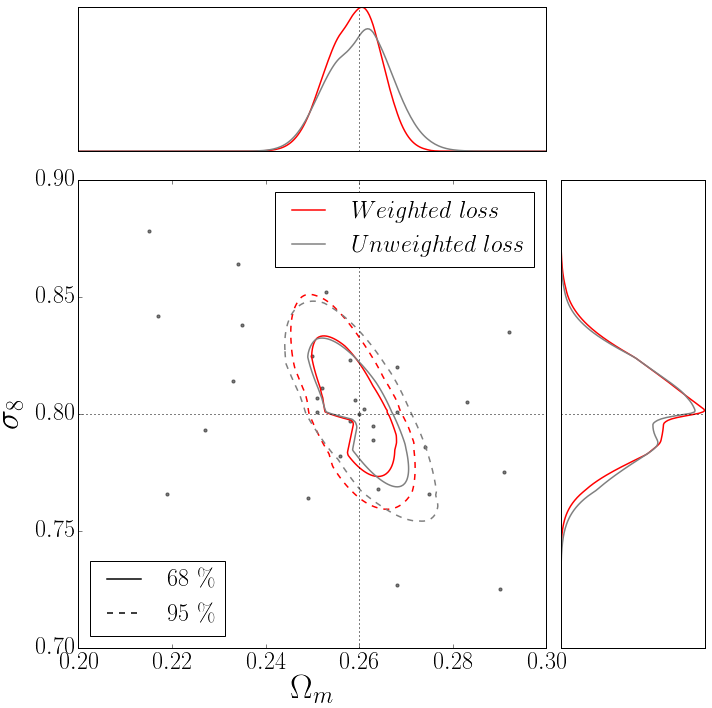}
\end{center}
\caption{\it 68\% and 95\% credible contours for un-smoothed ($\approx 0.2$ arcmin/pixel) $\kappa$ maps, derived from two neural networks with the same architecture: in red the result from training with the weighed loss function (eq.~\ref{eq:weighted-loss-function}) and in grey the result from training with the un-weighted loss function (eq.~\ref{eq:unweighted-loss-function}). True values are indicated by black, dotted lines. The upper and right panels show the marginal distribution for $\Omega_m$ and $\sigma_8$, respectively.} \label{ml_bias_2}
\end{figure}

The parameter constraints for an observation near the fiducial model are not affected by the use of an unweighted loss function, as Fig.~\ref{ml_bias_2} illustrates. This is because the scatter of the predictions in densely populated areas does not increase significantly when the bias in the sparsely sampled areas is reduced with a modified loss function. We did not re-train our networks with a weighted loss function due to the additional computational cost, since the constraints from the network's predictions are essentially unchanged.

\section{Conclusions}\label{ml_conclusions}

We trained a convolutional neural network on simulated, noiseless, weak lensing convergence maps. We demonstrated that neural networks can outperform methods based on traditional observables such as the power spectrum, or even statistics previously shown to extract non-Gaussian information, such as lensing peaks. On data smoothed at 1 arcmin scales, within reach of upcoming surveys, the neural network outperformed the power spectrum by a factor of $\approx 5$ and the lensing peaks by a factor of $\approx 4$ (using the area of the confidence contour in the  \{$\Omega_m,\sigma_8$\} plane as a figure-of-merit).

We performed null tests to verify that the improvement in the parameter constraints reflect the network's ability to extract additional information present in the WL data, and is not a numerical artifact (for instance, some form of overfitting). This sets a lower limit to the cosmological information encoded in noiseless lensing maps, whether this is also the case in more realistic, noisy data sets, remains an open question. The network's constraints are limited by both the precision and bias of its predictions. Whether further improvements are reachable through a different network architecture, or a richer training data set, remains an open question and calls for further investigation.

Our results are consistent with previous findings in \cite{Ravanbakhsh16} for the 3D matter power spectrum and in \cite{Schmelzle17} for the ability of neural networks to distinguish WL data generated from different cosmologies. Some of the questions that future work will address are:

\begin{itemize}

\item \emph{Effect of noise on predictive power}. The presence of realistic levels of noise (e.g. shape noise) can pose challenges to neural network training \cite{Schmelzle17}. It remains to be shown if the $\approx 5 \times$ improvement in parameter constraints compared with the power spectrum is achievable with noisy data.

\item \emph{Propagation of systematics on constraints from neural networks}. Before neural networks can be used to infer parameters from weak lensing data, we need to understand the effect of the systematics present in the data on the resulting parameter constraints.

\item \emph{Scaling with survey area}. Since neural networks' training time steeply increases with the map size, it is important to assess how the constraining power from their predictions scale with map size, and how the scaling compares with that for alternative methods such as the power spectrum.

\item \emph{Network analysis}. While the interpretation of feature maps from deep networks (see \cite{Schmelzle17}) is not straightforward, it may provide valuable insights to design new summary statistics capable of extracting cosmological information from lensing observations.

\item \emph{Improvements in the network's training and architecture}. An extended exploration of training parameters (density of models in parameter space, number of independent examples per model, loss function, etc.) and architecture's features (convolutional kernel size, number of layers, etc.) will elucidate the effect of these choices in the resulting constraints.

\end{itemize}

\section*{Acknowledgments}
We thank Martin Kilbinger and the anonymous referee for valuable comments on the manuscript. The simulations to generate the data, and the training of the neural networks were performed at the NSF XSEDE facility, supported by grant number ACI-1053575 and the Habanero computing cluster at Columbia University. This work was supported in part by NSF Grant No. AST-1210877 and by the Research Opportunities and Approaches to Data Science (ROADS) program at the Institute for Data Sciences and Engineering (IDSE) at Columbia University. DH acknowledges support from a Sloan Research Fellowship. ZH gratefully acknowledges sabbatical support by the Simons Fellowship in Theoretical Physics and thanks New York University, where some of this work was performed, for their hospitality.

\appendix
\section{Gaussian likelihood approximation}
\label{ml_appendixA}
One way to assess how valid the Gaussian approximation is for the likelihood of a given observable is to estimate its probability density function (PDF) from our simulations without assuming any specific functional form. A non-parametric method to do that estimation is the Kernel Density Estimator (KDE). The main challenge to apply this approach to lensing peaks is how to achieve a density estimator in a high dimensional space with a limited number of independent vectors (512 per model). 

We performed an analysis with noisy data within the framework of a different study, that supports that a Gaussian likelihood is not a bad approximation for lensing peaks. The dataset corresponds to the same cosmologies used for this study, and the convergence maps have been smoothed with a characteristic scale of 1 arcmin, but they also have an ellipticity noise of $\sigma_{\epsilon} = 0.4$ present. To reduce the dimensionality of the observable, we performed an Independent Component Analysis (ICA) \cite{Herault85, Comon94}. This method provides the directions that maximize negative entropy, which can be interpreted as the directions in which the data is less Gaussian. As a pre-processing step, we whitened the data (i.e. we removed its mean and normalized its covariance), and then we projected the whitened data into the 9 directions found following ICA. We then used a KDE to estimate the PDF of the resulting data. While we found some non-Gaussianities, specially for peak counts corresponding to high significance, the effect on the likelihood (and corresponding credible contours) is limited. 

\begin{figure}
\begin{center}
\includegraphics[width=0.45 \textwidth]{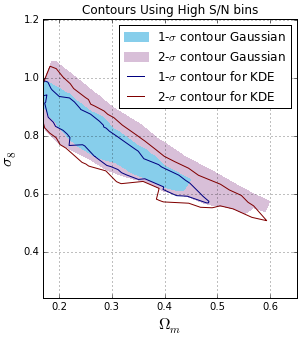}
\end{center}
\caption{\it Credible contours for $\{\Omega_m,\sigma_8,\Sigma_8\}$ from lensing peak counts on noisy $\kappa$ maps. Filled contours correspond to a Gaussian likelihood, and solid lines to contours corresponding to KDE estimates.} \label{ml_gaussian_peaks}
\end{figure}

As an illustration, in Fig.~\ref{ml_gaussian_peaks} we show the difference in credible contours obtained from a Gaussian likelihood from those obtained using a KDE. We display only the contours derived using only peaks with a signal-to-noise greater than 3. These are the peaks for which the non-Gaussianities are the most pronounced, and yet the contours obtained with both methods are comparable. Using a model to predict peak counts that does not rely on N-body simulations, \cite{Lin15} also found that a Gaussian likelihood is a good approximation (to $\sim$10\%) for lensing peaks.

To analyze whether a Gaussian distribution is a good approximation for the $\{\Omega_m, \sigma_8\}$ predictions from the neural network we used a modification of the Kolmogorov-Smirnoff test that can be applied to two-dimensional distributions \cite{Fasano87}. For each model, we computed the mean and covariance from the predictions for the test maps. Then, we tested the predictions against a Gaussian distribution defined by the estimated mean and covariance. 

The null hypothesis, that there is no statistical difference between the distribution of our empirical samples (neural network predictions) and a Gaussian, cannot be rejected with a confidence of 99\% except for 2 models which are far from the fiducial, $\{\Omega_m=0.450, \sigma_8=0.200\}$ and $\{\Omega_m=0.452, \sigma_8=0.454\}$. We conclude that a Gaussian likelihood is a reasonable approximation for the predictions from the neural network.

\section{Sensitivity of results to interpolation}
\label{ml_appendixB}
To assess how sensitive our results were to the models sampled from the parameter space $\{\Omega_m, \sigma_8\}$, we trained an additional network on the same un-smoothed $\kappa$ maps but removing the model $\{\Omega_m=0.261,\sigma_8=0.802\}$ from the training data set. When fed the test maps for that cosmology, the network that was not exposed to it during training yielded somewhat different predictions than the network which had seen maps from that model during training. The differences in the mean prediction were very small, with a shift of $-1.0\%$ in $\Omega_m$ and $-0.1\%$ in $\sigma_8$. The change in scatter is more significant, the standard deviation in the predictions for $\Omega_m$ increasing by $80.8\%$ and that for the $\sigma_8$ predictions by $12.2\%$. The larger degradation for $\Omega_m$ may be related with the fact that the network's architecture seems to have greater difficulty in distinguishing between models that differ in that parameter, as was shown in \S\ref{ml_discussion} for both GRFs and smoothed convergence maps.

\begin{figure}
\begin{center}
\includegraphics[width=0.5\textwidth]{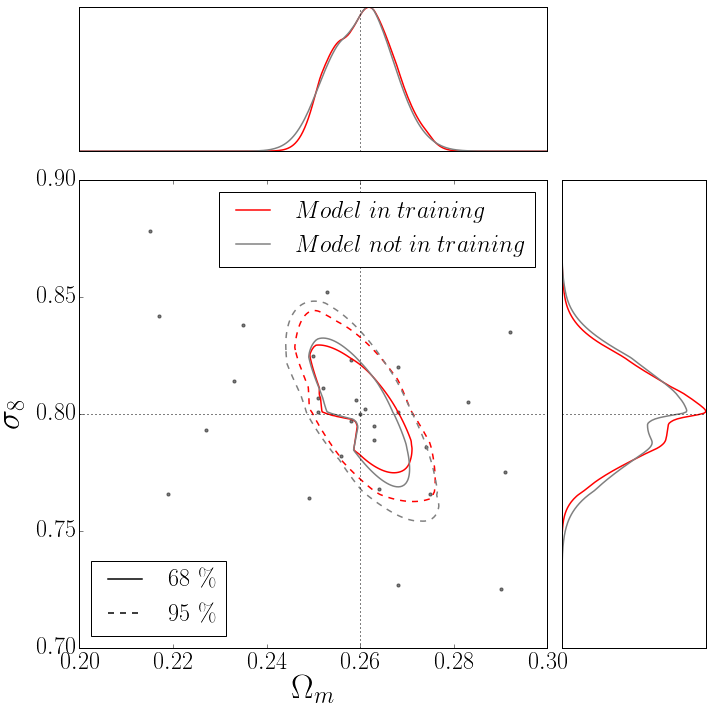}
\end{center}
\caption{\it 68\% and 95\% credible contours for un-smoothed ($\approx 0.2$ arcmin/pixel) $\kappa$ maps, derived from two neural networks with the same architecture: the original one trained on all 96 cosmologies (red) and another one for which the model $\{\Omega_m=0.261,\sigma_8=0.802\}$ was excluded (grey). The assumed true value ($\{\Omega_m=0.261,\sigma_8=0.802\}$) is indicated by black dotted lines. The upper and right panels show the marginal distribution for $\Omega_m$ and $\sigma_8$, respectively.} \label{ml_contours_interpolation}
\end{figure}

While this sensitivity to interpolation highlights how relevant a well-sampled training data set is for proper generalization by the network's architecture, we are mostly concerned about how interpolation errors propagate into the inferred parameters' constraints. That effect is small, as Fig.~\ref{ml_contours_interpolation} shows. The credible contours inferred from the predictions by both networks barely change, and the same applies to the marginal distributions inferred for both $\Omega_m$ and $\sigma_8$. We show the contours computed for the worst-case scenario, that is, when the model missing from the training data-set is the ``true" cosmology.

The small change in the parameter constraints' from both networks indicate that our main conclusions would not change with a different sampling of the parameter space. Besides, as the priors on our cosmological parameters improve with new experiments, the parameter volume to be explored will shrink and the number of models that need to be simulated to sample that space properly will also decrease.

\bibliography{ref}

\end{document}